\documentclass[prb,preprint]{revtex4-1} 
\usepackage{amsmath}  
\usepackage{amsfonts} 
\usepackage{graphicx} 

\newcommand{\be}{\begin{equation}}
\newcommand{\ee}{\end{equation}}
\newcommand{\bea}{\begin{eqnarray}}
\newcommand{\eea}{\end{eqnarray}}

\begin{document}

\title{A model for including Arduino microcontroller programming in the introductory physics lab}

\author{Andrew J. Haugen}
\affiliation{Department of Physics, Winona State University, Winona, MN 55987}

\author{Nathan T. Moore}
\email{nmoore@winona.edu} 
\affiliation{Department of Physics, Winona State University, Winona, MN 55987}

\date{\today}
\begin{abstract}
The paper describes a curricular framework for introducing microcontroller programming in the University Physics lab.  The approach makes use of Modeling Instruction, an effective approach for teaching science at the secondary level in which student learn the standard material by developing and deploying models of the physical world. In our approach, students engage with a context-rich problem that can be solved with one or more sensors and a microcontroller. The solution path we describe then consists of developing a mathematical model for how the sensors' data can be mapped to a meaningful measurement, and further, developing an algorithmic model that will be implemented in the microcontroller.  Once the system is developed and implemented, students are given an array of similar problems in which they can deploy their data collection system. Results from the implementation of this idea, in two University Physics sections, using Arduino microcontrollers, are also described.
\end{abstract}

\maketitle 

\section{Introduction}

When you ask a faculty member why they decided to study one of the hard sciences, the answer always seems to be some variant of ``I found it interesting.''  Saving the world and landing a job are nice perks, but as Feynman said, 
``The practical results aren't why we do it."

Mark Frauenfelder, the editor of MAKE magazine, explained this on the satirical ``Colbert Report" in 2010\cite{make_colbert}.  In the interview Frauenfelder showed the host, Colbert, some wooden spoons he'd carved from a tree branch that had fallen during a storm.  During the interview Colbert intimated that the spoons were valuable as gifts because they were in some sense a part of the person who'd made them.  Implicitly, the value of the spoon wasn't only in its function, but also in its creation.
The knowledge students discover and create for themselves when studying science is similarly precious and special.  A corollary may be that this is why student misconceptions are so difficult to dislodge - they are part of a student's self-created story of the world.   

Given the current zeitgeist surrounding the Arduino microcontroller, \cite{arduino}, it seems compelling to include a programming and sensor development component to the introductory labs, as it may increase the sense of wonder and ownership students bring to their work in that class.  At the same time, it seems that Modeling Instruction, \cite{modeling_cycle_1,modeling_cycle_2}, provides a valuable conceptual framework for the development of data acquisition equipment and microcontroller programming.  This paper describes a curricular framework for including microcontroller programming in the introductory lab, and describes initial results from piloting this idea in two sections of University Physics at Winona State, in 2011 and 2013.  

\section{The Arduino}

The Arduino microcontroller was initially developed in around 2005 as a platform for design students in Italy to learn how to develop interactive art exhibits.
The ``standard" Arduino configuration is updated regularly;  circa 2014, the 
``Arduino UNO" is based on the ATMega 328 chip, an 8-bit, 16MHz CPU with 2KB ram for program execution, and 32KB flash ram for program storage.    
The unit has 20 pins of digital IO. Several of these pins can also be used as analog inputs to measure voltages corresponding to temperature, humidity, pressure, distance, etc.
Units are available from a variety of vendors online for about $\$30$ per board. 
The ATMega CPU, which the Arduino is built around, is commonly used for automation and control tasks, for example, running a home thermostat. 

The standard way to program an Arduino is via C++, compiled in the Arduino integrated development environment (IDE), \cite{arduino_programmer}.  Alternatively, National Instrument's Labview programming environment can use the Arduino as a ``slave" data acquisition device, \cite{labview_programmer}.   The initial work described has implemented trial curriculums using both of these programming paradigms.

Further, Arduino boards can communicate with a computer via USB using a virtual serial port. Nearly any programming language is able to communicate through a serial port, \cite{interfacing}, and so interfaces in Matlab, Mathematica, Python, PERL, etc are also available.  

The board is widely adopted by hobbyists and many examples are available online. Google gives about 88 million hits for ``Arduino Programming" in July 2014, and a number of hobbyist-level books on the system have been published. Notably, a few physics faculty members have implemented lab activities with an Arduino, \cite{quarkstream, new-york-arduino-book}, but so far, it does not seem that student work with the device has been included wholesale in introductory classes.

\section{Conceptual framework for  lab development}

In the work described, lab periods were modified in the following way.  At the beginning of the semester all students bought a $\approx\$60$ lab kit consisting of an Arduino and set of sensors. Over the course of the term, in roughly alternate weeks, students first developed microcontroller-driven sensor systems to perform a given measurement, and then deployed their systems to address a set of similar problems.  The problems developed related to the normal topics in University Physics, but the approach in lab was quite unconventional, and  required a few curricular components which may seem novel to a traditional lab class.  

First, as most of the sensors used interface with the Arduino via a voltage, the curriculum needs to begin with a simple discussion of electronics, e.g. voltages, currents, soldering, avoiding wool pants in the lab, etc.

Second, it has been our experience that nearly all of the students in University Physics arrive at the University with no programming background, and the basics of computational/algorithmic thinking will need to be introduced.
Others, \cite{Chabay_Sherwood}, have shown both the utility and feasibility of including computational thinking as part of an introductory physics sequence.
The necessary programming concepts: variables, control structures, recursion, and IO need to be introduced at the beginning of the course and then will be enriched and further developed along the way.  The simple ``Hour of Code,'' \cite{hour_of_code} exercise and a follow-up discussion might be sufficient for this introduction. 

Within lab problems it seems appropriate to use a modified version of the ``Modeling Cycle," \cite{modeling_cycle_1,modeling_cycle_2},  adapted to data-acquisition (DAQ) system development.  Specifically, lab activities are structured in terms of the following stages:
\begin{enumerate}
\item Context-Rich Concrete Preparation
\item Sensor Data Collection
\item Mathematical Model-Building
\item Algorithmic Model-Building
\item Model Deployment
\end{enumerate}
In each stage, students work in small, meaningfully chosen groups.  At various stages in the lab process, the class breaks for short ``whiteboarding" sessions, \cite{whiteboarding}, during which groups describe to each other what they've figured out thus far.   

A brief summary of the purpose and context of each stage follows.  As an appendix,  a sample lab activity using the LM34 Temperature probe is included.  

\textbf{Context-Rich Concrete Preparation}
Following the work of others, \cite{Shayer_and_Adey,Heller}, the activity begins with common language and a compelling problem to be addressed by the students.  Given that the approach to lab will seem unusual, and perhaps like extra work, the activity will need to ``feel" like something an actual scientist or engineer will do in their daily activities.  Part of this context can be provided before the lab. This preparation probably includes some problem specifications and data sheets for relevant sensors. 

\textbf{Sensor Data Collection}
In teams, the students breadboard the sensors and perform basic calibration tasks: How does voltage map to length? To what standard deviation are measurements repeatable? etc. 
Student inquiry in this stage is guided by meaningful questions that must be answered by actually working with the hardware (i.e., they likely cannot be answered by simply reading the specification sheet for a given component).  For example, one might ask how a temperature probe responds to rising vapor above a beaker of boiling water -- details like this are rarely included in data sheets.  

\textbf{Mathematical Model-Building}
With basic data gathered, students build mathematical models which map sensor readings (voltage, current or similar) to physical parameters that the sensors are reading (sound pressure, light intensity, acceleration, etc). This is more than simple calibration, as the students are expected to create (not just locate and regurgitate) models that relate sensor outputs to physical measurements.  For more on mathematical model-building, see \cite{modeling_cycle_1,modeling_cycle_2}.

\textbf{Algorithmic Model-Building}
With a mathematical model developed for the sensors, the students then create an algorithm that can be implemented in either an Arduino or in a Labview program.  These short programs solve their initial problem.   Writing an algorithm to control the Arduino needs to be separate from building a model of the sensor as these are distinct model-building tasks.  

\textbf{Model Deployment}
Finally, with the basic instrument and interface developed, the students tackle (in lab, or as homework) more sophisticated permutations of the original design problem.   

Structuring the labs in this way should develop a constructivist pattern that students should  be able to adapt to new sensors and new algorithm demands.  In principle, this is a learning pattern that should transfer beyond the context of University Physics.

\section{Results and Conclusions}
The lab structure described is the result of two semesters of trial implementation of this DIY style approach to DAQ development in the introductory labs.  

In the Spring 2011 section of University Physics 1 all of the students bought a ``labkit" consisting of an Arduino, an ultrasonic distance sensor, and other equipment which allowed students to write data to an SD flash card.  Students programmed their Arduinos in the text-based Arduino language via the standard IDE.  Lab experiments included calibration of a temperature probe, work with the ultrasonic distance sensor, and various position measurements via a GPS antenna.  

The Spring 2013 section was similar, although this time students controlled and interfaced to the Arduino via Labview programs written by students on their laptops.  Labs were similar, also including acceleration and $\vec{g}$ measurements via an 3-axis accelerometer.  

This approach to lab work seems to be novel, and none of the standard Physics Education pre/post inventories appropriately mapped to the modified curriculum.  When measured for these sections, MPEX, FCI, CTSR, etc scores were all unremarkable from standard sections of the class.  

Anecdotally, in each iteration there were a handful of students who latched onto the idea and produced remarkable results.  For example, during the 2011 semester one student built a simulation of the Ranger 7 mission to the moon, \cite{Ranger_7}, on his Arduino board (rather than via a spreadsheet, as the rest of the class was instructed).  Later on, some students in the 2011 section went on to use Arduino boards in their senior engineering design projects.  Similarly, some of the 2013 students took National Instrument's initial Labview Certification Exam, \cite{labview_certification_exam}, and at least one student landed a summer internship in part because of his exposure to Labview interfacing in the introductory course.  

In both semesters however, we underestimated the extent to which this idea needed to be ``sold."  In our eyes, this early exposure to real interfacing, programming, and engineering was incredibly valuable, but to the students it sometimes came across as extra work, somewhat tangential to both  the course syllabus and their future careers.  

In future semesters it will be essential to both remind students of the long-term value of engaging in real design work, and also modify semester exams to better incorporate the modifications we make to our labs.

Gratefully thanks goes to Tia Troy, Megan Reiner, Scott Stroh, and Tucker Besel for their help in implementing this idea as learning assistants.  Thanks also goes to students in the Spring 2011 and 2013 sections of Physics 221 for their participation in the project.    

\appendix

\section{Sample Lab Structure for an energy activity using the LM34 Temperature Probe}

\subsection{Background}
Over the summer you have a job working for a medical technology start-up in Rochester, MN.  The company is working on a high-precision infra-red incubator to be used for newborn babies in the maternity ward of hospitals, and you're excited to be able to help on the project.

\begin{figure}[h]
\begin{center}
\centering
\includegraphics[width=0.4\textwidth]{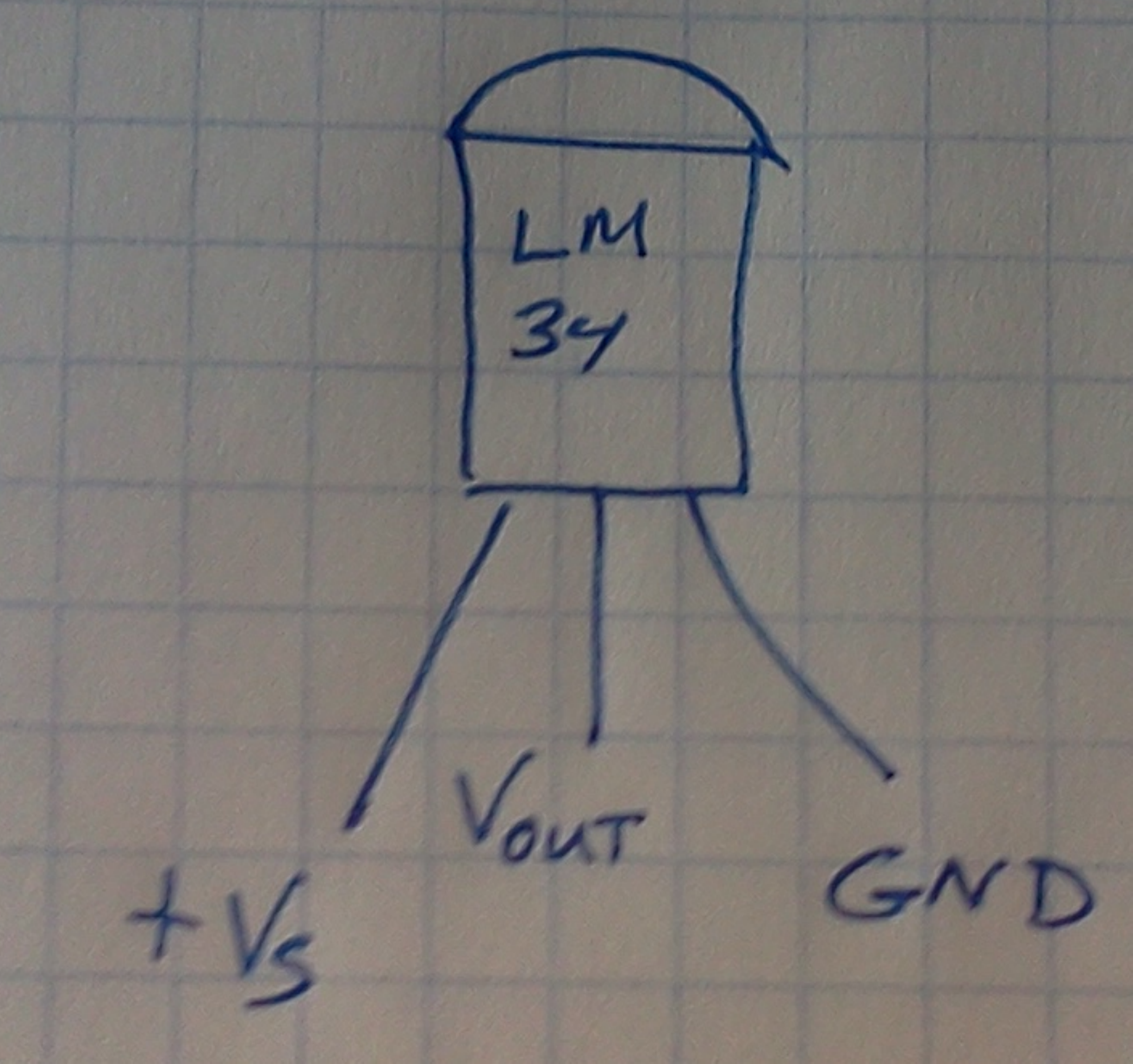}
\label{fig:sketch}
\caption{Wiring pin-out for an LM-34 Temperature probe.}
\end{center}
\end{figure}

In your first week, your boss asks you to familiarize yourself with temperature measuring equipment (obviously, the incubator can't allow the baby to be chilled or overheat, and the control circuitry has to be designed accordingly).  The specific device she wants you to work with is an LM34 temperate probe, which is made by National Semiconductor (among others).  A data sheet for the LM34 is available, \cite{LM34_spec_sheet}.  Please read through the datasheet.

The specific wiring diagram for the LM34 is given in figure \ref{fig:sketch}.  From reading through the datasheet, you should be able to determine appropriate values for $V_S$ and the numerical conversion from $V_{OUT}$ to a sensor temperature, measured in degrees Fahrenheit.

You should be able to supply $V_S$ and $GND$ from your Arduino, and $V_{OUT}$ can be read either with a voltmeter or one of the analog-to-digital (A2D) pins on your Arduino.

\subsection{Data Collection}
After wiring up your LM34 sensor, please take data (using a glass thermometer as a control), which characterizes the behavior of the sensor.  Please make sure your data allows you to answer the following questions:
\begin{enumerate}
\item What pin, and in what way do you measure sensor voltage?
\item How do you convert the sensor voltage into a sensor temperature?
\item How long does it take the sensor to equilibrate with its surroundings?
\item What's the min/max surface temperature of your body? (be modest!)
\item How does the value read from the LM34 compare (in accuracy) to a reading from the standard glass thermometer?
\item Is the calibration formula given in the datasheet reliable?  
\item To what accuracy does the sensor allow you to measure temperature?
\end{enumerate}

\subsection{Building a model}
With data collected, please build a mathematical model which allows you to predict sensor temperature based on the voltage the sensor reports.  To create this model, you should first produce a graph of sensor temperature as a function of output voltage.  A mathematical model should be accessible from the data you plot.

\subsection{Building an Algorithm}
From what you already know about Arduinos, please create an algorithm which illustrates the following procedure.  
\begin{enumerate}
\item The sensor is in contact with an object whose temperature we wish to measure.
\item The sensor outputs a voltage which corresponds to a temperature.
\item The Arduino reads this voltage via an A2D pin.
\item The Arduino converts the integer A2D value into a temperature, in Fahrenheit (using your model, above).
\item The Arduino sends the temperature reading and a time measurement to the serial monitor (e.g., your laptop).
\item The Arduino waits a specified time, and then repeats the sequence of procedures.
\end{enumerate}

Once you have an algorithm sketched out, please build and test code which executes it.

\subsection{Deploying the model}
Finally, please work on one of the following design problems which make use of your model and algorithm.  Your lab instructor will assign the specific problem for you to work on.

\begin{enumerate}
\item In the infant incubator, the IR heating lab should turn on if the sensor reads an infant finger temperature lower than $T_1=98.0^{\circ}F$, and further, the heating lamp should turn off when the temperature exceeds $T_2=100.0^{\circ}F$.  Using your model and initial algorithm, create and implement a new algorithm which executes this temperature-control procedure.  You can represent the heater's function via an LED, which would be driven by the Arduino.  In an actual design, the Arduino would drive a relay or transistor, which controls the heater.

\item Children are precious, and it won't do to trust the life of a child to a single $\$0.76$ component.  For greater safety and reliability, consider an improved design which would read in values from two LM34 sensors and actuate a heater if and only if both sensors agree.  Create and implement such an algorithm.

\item A rice cooker has two heater settings, the first to get the rice+water up to boiling, and a second heater setting to keep the rice warm once it has fully cooked.  Design and implement an algorithm to model the function of a rice-cooker.  You can use LED's to represent the function of the two heaters.

\item Whenever you open a refrigerator door, cold air leaks out and the compressor has to run to cool down your food.  Create and implement a data collection algorithm that would allow you to capture this data to see how much time it takes to cool the fridge down to proper temperature after the door has been opened. 

\item In general, the mathematical equation describing the temperature, $T$, of an object in contact with a larger heat reservoir with temperature, $T_H$, is given by 
\be \frac{dT}{dt}=-\kappa (T-T_H).\ee
Solve this equation, and compute $\kappa$ for the conditions that your fingers are the ``infinite heat reservoir" that the LM34 is in contact with. What implications does this time have in the algorithmic/mathematical models you built earlier in the lab?

\end{enumerate}

\end{document}